\documentclass[prd,onecolumn,12pt]{revtex4}
\usepackage{amsmath}
\usepackage{color,graphicx}

\begin{document}

\title{Perturbative QCD Analysis of Local Duality in a fixed $W^2$ Framework.}

\author{S.~Liuti,$^{a}$}
\email[]{sl4y@virginia.edu}
\author{R.~Ent,$^{b}$} 
\email[]{ent@jlab.org}
\author{C.E.~Keppel,$^{b,c}$} 
\email[]{keppel@jlab.org}
\author{I.~Niculescu$^{b}$}
\email[]{ioana@jlab.org}

\affiliation{
$^{a}$ University of Virginia, Charlottesville, Virginia 22901
\\
$^{b}$ Thomas Jefferson National Accelerator Facility, Newport News, 
Virginia 23606
\\
$^{c}$ Hampton University, Hampton, Virginia 23668}
                                                       
\date{\today}

\begin{abstract}
We study the global $Q^2$ dependence of large $x$ $F_2$ nucleon structure 
function data, with the aim of providing a perturbative-QCD (pQCD) based, 
quantitative analysis of parton-hadron duality. As opposed to previous 
analyses at fixed $x$, we use a framework in fixed $W^2$. 
We uncover a breakdown of the twist-4 approximation with a 
renormalon type improvement at ${\cal O}(1/Q^4)$  
which, by affecting the initial evolution of parton distributions, will have
consequences for pQCD analyses also at large 
$x$ and very large $Q^2$.
\end{abstract}

\maketitle


One of the key challenges in Quantum ChromoDynamics (QCD) today is to
formulate a connection 
between the description of the hard, or short-distance, scattering processes, 
which can be 
calculated in terms
of quark and gluon degrees of freedom
using perturbative methods, and  
the physical asymptotic states,  
{\it i.e.} the spectra of hadrons which
are not calculable within perturbative-QCD (pQCD) 
and are in principle 
only remotely related to parton dynamics \cite{Dok}. Yet,
a substantial number of observations indicate that 
QCD has manifestly a {\em dual} parton-hadron nature. 
Recent investigations of hadronic jets 
at $e^+e^-$ \cite{DELPHI}, $ep$ \cite{zeus00}, and $p\bar{p}$ \cite{CDF} colliders
show that several infrared-sensitive features of the inclusive hadron distributions 
can be reproduced by a pQCD description of the parton shower down to 
$Q_o \approx \Lambda_{QCD}$, $Q_o$ being an effective cut-off 
defining the onset of the non-perturbative regime
and $\Lambda_{QCD}$ being the scale of QCD.
(Notice, however, that detailed angular 
correlation measurements have been found 
to be in disagreement with available pQCD estimates \cite{malcolm}.) 
Similarly, deep inelastic scattering (DIS) experiments at very low
Bjorken $x$ indicate that pQCD can describe the nucleon 
structure function $F_2$ 
down to $Q^2  \approx 1 \, {\rm GeV^2}$, where  
a smooth and fast transition to the non-perturbative regime occurs, 
leading eventually to the expected behaviour in the 
real photon scattering limit ($F_2(x,Q^2) \rightarrow 0$ for $Q^2 \rightarrow 0$).
The question of the role and nature of
non-perturbative corrections naturally emerges. These
are known to affect
the cross sections for $e^+e^-$ jet fragmentation \cite{ALEPH} 
and are also expected to be present, although they
have not been isolated yet, in the deep inelastic data at 
low $Q^2$ and low $x$ \cite{Golec}.

In this Letter we address yet another 
set of experiments, 
namely inclusive $ep$ and $eD$ scattering in the resonance region,
where a pQCD description seems to hold in spite of the low values
of the invariant mass squared, $W^2$, produced in the final state. 
Here, the observation of Bloom-Gilman (BG) duality \cite{BG}, or
the similarity between the  
behaviour of the resonance contribution to the nucleon structure 
function and DIS, 
can be formulated theoretically, as the equivalence between the moments 
of the structure function in the low $W^2$ kinematical region dominated 
by resonances and in the DIS one, {\em modulo} 
perturbative corrections and expectedly small power corrections \cite{DGP}.
Furthermore, 
because of the large body of highly accurate data that has been recently made 
available \cite{ioana1}, the data in the resonance region 
can now be fitted 
to a smooth curve which, once evolved according to pQCD, will coincide with
the DIS data. 

There has been a growing theoretical interest in this intriguing phenomenon, 
where the quasi-exclusive nature, or low inelasticity of the scattering process  
would be expected to hinder a direct observation of the partonic structure 
of the target. On the contrary, the measured structure functions appear
to behave, in average, like parton distributions.     
The present analysis is motivated by the expectation that a more precise 
understanding of the mechanisms behind BG duality 
might provide a handle on the type of hadronic configurations 
that are present in a ``semi-hard'' regime, before confinement settles in.   
It is therefore of paramount importance to perform a detailed study of both 
the logarithmic corrections and of the size and nature 
of the power corrections in the pQCD expansion, in order to ascertain whether the apparent 
weak $Q^2$ dependence of the data is coincidental, an artifact of the particular 
region under study, or a cancellation of Higher Twist (HT)
terms, possibly understandable within parton-hadron duality models.     
The analysis conducted here involves a number of steps similar to recent 
extractions of power corrections from inclusive data 
\cite{YanBod,AKandco,ScScSt,twistpaper}, namely the form 
\begin{equation}
F_2^{exp}(x,Q^2) = F_2^{pQCD+TMC}(x,Q^2) + \frac{H(x,Q^2)}{Q^2} +{\cal O}(1/Q^4),   
\label{t-expansion}
\end{equation}
is adopted, where $F_2^{pQCD+TMC}(x,Q^2)$ is the twist-2 contribution, including  
kinematical power corrections from the target mass (TMC); the other terms in the 
formula are the dynamical power corrections, formally arising 
from higher order terms in the twist expansion. 
Both $F_2^{pQCD+TMC}$ and $H$ can be extracted from the data
at large $x$, by taking care of aspects of pQCD evolution peculiar to this region,
and of TMC. Uncertainties are introduced at each step. 
In this Letter, we address them one by one with the important addition 
that we provide for the first time
in the literature, both an analysis of the scale dependence of the resonance 
region ($W^2 < 4 \, {\rm GeV}^2$) and a combined study of both the resonance
and the large $W^2$ data.    

We start by studying the question 
of the limit of applicability of pQCD in the low $W^2$, large $x$ domain, but 
at $Q^2$ above the $1$ GeV$^2$ scale, so still amenable, in principle, 
to a pQCD treatment.
Previous
analyses used moments of the structure functions obtained at fixed values 
of $Q^2$ and directly related to the OPE, circumventing the problem of 
dealing with the complicated structure of the bound states by providing a 
natural averaging procedure 
in the Mellin conjugate $n$
\cite{DGP}.
However, we have shown in \cite{Entetal} that, at lower values of $Q^2$, such
moments of $F_2$ are affected by elastic scattering in such a way
as to render higher twist contributions impossible
to extract or interpret in the standard OPE language. We choose, therefore,
an alternative method using data in fixed $W^2$
bins.

The data in the resonance region \cite{ioana1} have been fitted to   
a smooth curve for $F_2^{p(D)}(\xi)$,  
where $\xi= 2x/(1+\root \of {1+4M^2x^2/Q^2}$) \cite{nacht} is the Nachtmann
scaling variable.
The fit is applied separately to bins in invariant mass, $W^2$, centered at: 
{\bf (1)} $W_R^2=1.6$ GeV$^2$, {\bf (2)} $W_R^2=2.3$ GeV$^2$,  
{\bf (3)} $W_R^2=2.8$ GeV$^2$, {\bf (4)} $W_R^2=3.4$ GeV$^2$, respectively. 
The $\chi^2 /d.f.$ for these fits varies from 0.8 to 1.1. The
uncertainty for the scaling curves is estimated to be
better than 10\%, taking into account uncertainties in the experimental data,
and in the averaging and fitting procedures.
The curves and uncertainties in each $W^2$ bin 
are represented by the hatched areas in Fig. 1.
Notice that the spectra at fixed $W^2$ require that  
$Q^2$ vary for each spectrum as an increasing function 
of $x$, since $Q^2 \equiv Q^2(x) = (W_R^2 - M^2) x/(1-x)$.


We utilize a quantitative comparison of the resonance region $F_2$ curves
thus obtained to the pQCD prediction for $F_2$ at the same kinematics to 
extract potential higher twist contributions to the structure functions.
We assume that only valence quarks contribute 
to $F_2^{p(D)}$, at low $W^2 \lesssim  10 \, {\rm GeV^2}$ 
and $x \geq 0.3$.
PQCD evolution to $Q^2 \equiv Q^2(x)$ or at fixed $W^2$, 
for the Non-Singlet (NS) distributions 
at next-to-leading-order (NLO), is given by: 
\begin{eqnarray}
q_i^{(-)}(x,Q^2) & = & \int_{Q_o^2}^{Q^2(x)} \frac{dQ^{\prime,2}}{Q^{\prime,2}}  
\frac{\alpha_S(Q^{\prime,2})}{2\pi}
\nonumber \\  \times & &
\int_x^1 \frac{dy}{y} P_{qq}\left( \frac{x}{y},
\alpha_S(Q^{\prime,2}) \right) q_i^{(-)}(y,Q^{\prime,2}),
\label{eq:DGLAP}
\end{eqnarray}
where $q_i^{(-)}=q_i-\bar{q}_i \equiv q_i^v$, $i=u,d$.
The expressions for the splitting function 
$P_{qq}\left(z,\alpha_S(Q^2) \right)$,
and for the corresponding 
endpoints at NLO in $\overline{MS}$ scheme can be found in \cite{CFP,HW};
$\alpha_S$ is the strong coupling constant.
The structure function, $F_2$, 
is then obtained by convoluting (\ref{eq:DGLAP}) with the quark coefficient function, 
$B_2^{NS}$ \cite{CFP}. 
We fix the values of the initial parton distribution functions (PDFs), at
$Q_o^2 \approx 0.4 - 1$ GeV$^2$,
to the ones taken from NLO global fits to world 
data \cite{GRV98,MRST,CTEQ5}, and we solve the evolution equations 
directly in $x$ space, with $\alpha_S(M_Z^2)=.117$. 
Notice in this respect, that the shape of the initial NS PDFs 
is practically constrained \cite{MRST}, at variance with
the singlet and gluon distributions at low $Q^2$, 
whose shape is strongly correlated with the value of $\alpha_S$.
The structure of perturbative evolution, including both coefficient functions
and splitting functions for the NS part, can now be evaluated up to NNLO.
Detailed studies of the impact of NNLO corrections and beyond, on the 
determination of power corrections for the NS structure functions,  
have been performed in \cite{AKandco,ScScSt}: The question of whether  
these can ``mimick'' the contributions of higher twists, including 
the uncertainties due to the well known scale/scheme dependence of 
calculations, within the current precision of data is under intense investigation.
Here, we single out the contributions that are expected to dominate the higher 
order perturbative predicitions at large $x$, namely powers of $\ln(1-z)$ 
terms, where $z=x/y$.  
These terms can be resummed to all orders. 
We perform the resummation in $x$ 
space by replacing the $Q^2$ scale 
with a $z$-dependent one, $\widetilde{W}^2 = Q^2(1-z)/z$ \cite{Bro}. 
It is well known that such a procedure introduces in principle 
an ambiguity in the evaluation of the  
running coupling at low values of the scale $\widetilde{W}^2$, {\it i.e.} 
as $z \rightarrow 1-\Lambda^2/Q^2$ 
(\cite{Rob_pap} and references therein).
This ambiguity is lessened in our analysis 
because at fixed $W^2$, $\Lambda^2/Q^2(x)$ is very close to zero.
Our results for NS evolution and shown elsewhere \cite{Liu}, are in good agreement with 
more recent resummation calculations performed 
in $n$ space and anti-Mellin transformed as in \cite{SteVog}. 

Finally, we take into account target mass corrections (TMC). 
We use the expression in \cite{DGP}, obtained by 
a formal inversion of the Nachtmann moments \cite{nacht}. 
Ambiguities in this procedure are expected to arise because 
of the neglect of higher order and higher twist corrections. 
To safely disregard these, we use a minimal criterion that only 
the kinematical 
points yielding low values of the parameter $x^2 M^2/Q^2$ in
the TMC expansion \cite{DGP} are kept. 

%
Comparisons of pQCD$+$TMC predictions with the resonance
average data are shown in Figures 1
and 2. 
A few comments are in order. In Fig. 1: {\it i)} TMC (full lines) modify 
substantially the pQCD behavior (dotted lines) rendering a better 
agreement with the data; {\it ii)} the
curve calculated using NLO pQCD at $Q^2=200$ GeV$^2$ 
shown for comparison in Fig. 1(b) demonstrates the large 
effect of pQCD corrections above $x \sim 0.2$.


In Fig. 2, 
we show the low $W^2$ data obtained from the fit of Jefferson Lab (JLab) 
and SLAC data in the resonance region, along with
larger $Q^2$ data from \cite{Dasu,BCDMS}. 
Note that
the data in the resonance region (circles) smoothly blend to the deep
inelastic (stars) - another manifestation of BG duality. The curves
correspond to our
calculations including pQCD+TMC at NLO (dashes), 
and pQCD+TMC with resummation (full).
The dots in each curve represent regions where TMC are 
uncertain. 
The effect we find is qualitatively similar to
what found in \cite{AKandco,ScScSt}, in that over the range 
$0.45 \leq x \leq 0.85$, higher order perturbative 
contributions, in this case large $x$ resummation, 
improve the agreement with the data.
Substantial discrepancies remain which
we assume to be largely due to 
the dynamical, HT corrections
to the structure function. 
If $H(x,Q^2)$ is modeled 
similarly to previous extractions \cite{VirMil,YanBod,AKandco,ScScSt}, one has:
\begin{equation}
H(x,Q^2) = F_2^{pQCD+TMC}(x,Q^2) C_{HT}(x),
\label{CHT}
\end{equation} 
Eq.(\ref{CHT}) is 
motivated by the lack of knowledge of the anomalous dimensions
of the twist-4 operators, a reasonable assumption
within the precision of the data (see also \cite{Ale}). 
Our analysis at fixed $W^2$ 
explained in the first part of the paper, enables us to
extract $C_{HT}$ from the resonance region
and from the large $W^2$ (DIS) region, separately. 


In Fig. 3(a) we show the coefficient $C_{HT}$, Eq.(\ref{CHT}), extracted from: 
{\it i)} DIS data with
$W^2 \ge 4$ GeV$^2$; {\it ii)} The resonance region, $W^2 < 4$ GeV$^2$; 
{\it iii)} Averaged over the entire range of $W^2$.     
The figure also shows the range of extractions previous to the current one 
\cite{VirMil,YanBod}.
We notice that in all three cases, our values for $C_{HT}$ are smaller than the 
ones in \cite{VirMil,YanBod}, because of the  
effect of large $x$ resummation. 
We have checked that our results without resummation 
are consistent with a previous extraction 
using moments of the structure function \cite{twistpaper}. 
Most importantly, while the large $W^2$ data track a curve that is 
consistent with the $1/W^2$ behavior expected from most models 
\cite{renorm}, the low $W^2$ data yield a much smaller 
value for $C_{HT}$ and they show a bend-over of
the slope vs. $x$, already predictable 
from a similar bend-over in the slopes at low $W^2$ in Fig. 2. 

Notice that this surprising effect is not a consequence 
of the interplay of higher order corrections
and the HT terms, but just  
of the extension of our detailed pQCD analysis to 
the large $x$, low $W^2$  kinematical region. 
In fact, resummation produces an overall reduction of $C_{HT}$, 
that does not modify our results qualitatively.         
In order to ascertain whether the discrepancy between the low $W^2$ 
and large $W^2$ values of $C_{HT}$ are due to ${\cal O}(1/Q^4)$ terms 
in the twist expansion, Eq.(\ref{t-expansion}), 
which could become more important 
at low $W^2$, we have extracted for each resonance
the quantity $\Delta H(x,Q^2)$, 
defined as
\begin{equation}
\frac{F_2^{exp}}{F_2^{pQCD+TMC}}= 1 + \frac{C_{HT}(x)}{Q^2} 
+  \Delta H(x,Q^2),
\label{DH}
\end{equation}
where $C_{HT}(x)$ coincides with the value fitted at large $W^2$.
From Fig. 3(b) one sees that 
$\Delta H(x,Q^2)$ is negative for all lower $W^2$ ($\le 3.4$ GeV$^2$)
bins, as expected if a cancellation among 
higher order inverse powers were to occur,
consistently with the requirement of parton-hadron duality. 
However, we uncover a non trivial $Q^2$ dependence of this term:
one can see a sharp change in the behavior of the 
higher mass resonances and  
the data in the $N-\Delta$ transition region which 
show a distinctively steeper fall with $Q^2$.
Furthermore, the high mass resonances  
seem to be in fair agreement with a simple 
$-1/W^4$ fit inspired by Infrared Renormalon (IRR) calculations 
\cite{renorm}, which would produce a constant line at a decreasing height 
for increasing $W^2$. The $N-\Delta$ transition region
shows a departure from this behavior which cannot be 
accounted for by fitting the first few terms of an inverse power expansion. 
A similar conclusion was reached in our analysis in moment space
\cite{Entetal}.
%

To summarise, the size of the power corrections obtained from the 
data at $0.45 \leq x \leq 0.85$, 
with $W^2 > 3.4$ GeV$^2$, is comparable to the 
values obtained in recent analyses of DIS, with 
$W^2 > 10$ GeV$^2$ \cite{VirMil,YanBod,Ale,AKandco,ScScSt}, 
the inclusion of large $x$ resummation producing a 
reduction of the coefficient $C_{HT}$. 
At lower invariant mass, $ 1.9 \leq W^2 \leq 3.4$ GeV$^2$,  
an extra term of order ${\cal O}(1/Q^4)$, with a negative 
coefficient is necessary to fit the data, in line with 
the asymptotic nature of the twist expansion 
and with current predictions from IRR calculations \cite{renorm}.    
A similar expansion does not reproduce, however, the scale dependence
of data at even lower masses, $1.2 \leq W^2 \leq 1.9$ GeV$^2$,
yet for $Q^2$ values where pQCD would be expected to apply. 
Therefore, the experimental observation 
of a rather flat $Q^2$ dependence cannot be taken as a signature
of parton-hadron duality for this region, but on the contrary, 
it shows the limits of applicability of this idea.     

Having this determination in hand, one can now attempt to provide 
theoretical interpretations. 
In particular, 
the fact that the pattern of dynamical power corrections 
to resonance production 
is comparable to the one for the large $Q^2$ deep inelastic
data, {\it i.e.} 
that power corrections remain small in the resonance region,  
suggests that color confinement is more likely to happen locally, 
with a smooth transition between 
partonic and hadronic configurations,  
a mechanism supported also by recent studies of hadron spectra in 
jet measurements \cite{CDF}.
This mechanism seems to break down in the $N-\Delta$ transition region, 
providing a threshold where cooperative effects from many partons
dominate the structure function. 
More information on the locality of color confinement 
may be obtained by studying  semi-inclusive 
processes where a more direct relation to cluster hadronization models 
can be established \cite{Liu}. 

We thank S. Alekhin for carefully reading the manuscript; W. Vogelsang for discussions 
and U.K. Yang for providing
us with details on his evaluation of higher twists. 
We also thank the Institute of Nuclear and Particle Physics 
at the University of Virginia and the National Institute for Theoretical 
Physics at the
University of Adelaide for hospitality and 
partial support during the completion of this paper.  
This work is supported in part by research grants from the U.S. Department
of Energy under grant no. DE-FG02-95ER40901 and the 
National Science Foundation under grants no. 
9633750 and 9600208 (Hampton).

\begin{figure} 
\includegraphics[width=15cm]{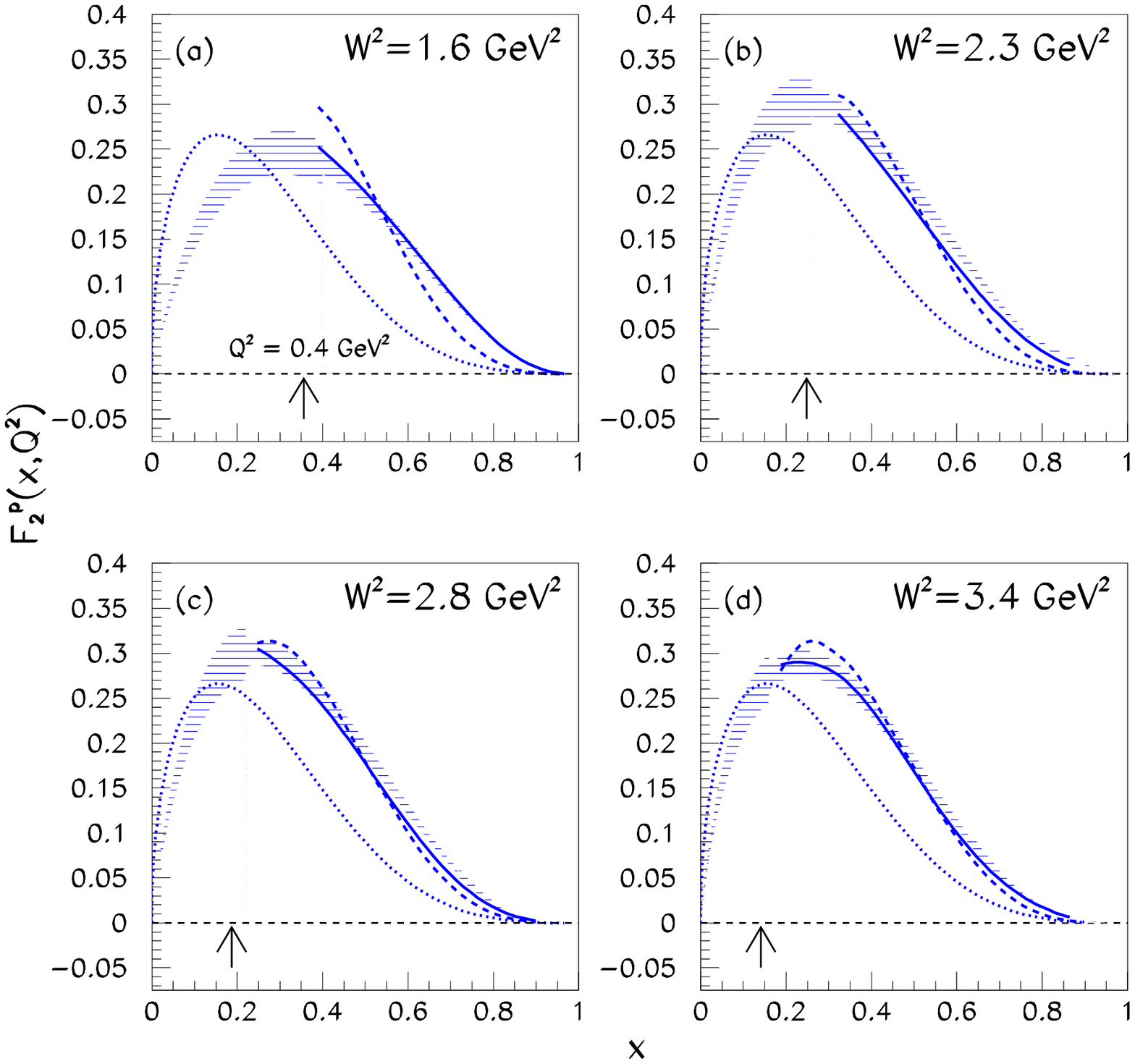}
\caption{Comparison of NLO pQCD calculations (dashed lines), and NLO pQCD+TMC (solid 
lines) with the data on $F_2^p$ (hatched areas) at fixed values of
$W^2=W^2_R$, vs. $x$: (a) $W_R^2=1.6$, (b) $W^2=2.3$, 
(c) $W^2=2.8$, (d) $W^2_R=3.4$ GeV$^2$. The dotted curve shows for 
comparison the DIS calculation, obtained at $Q^2=200$ GeV$^2$. The data 
are averaged with the procedure described in the text and reference. 
The pQCD curves were obtained using the GRV parametrizations for the 
NS distributions at the input value
of $Q_o^2= 0.4$ GeV$^2$ \cite{GRV98}, indicated by the 
arrows. Other parametrizations \cite{MRST} give similar results, starting from 
their input value of $Q_o^2\approx 1$ GeV$^2$. Below $Q^2=0.4$ GeV$^2$, 
the similarity between the pQCD curves and the 
data in the resonance region can no longer be tracked by pQCD.}

\end{figure}
\begin{figure}  
\includegraphics[width=15cm]{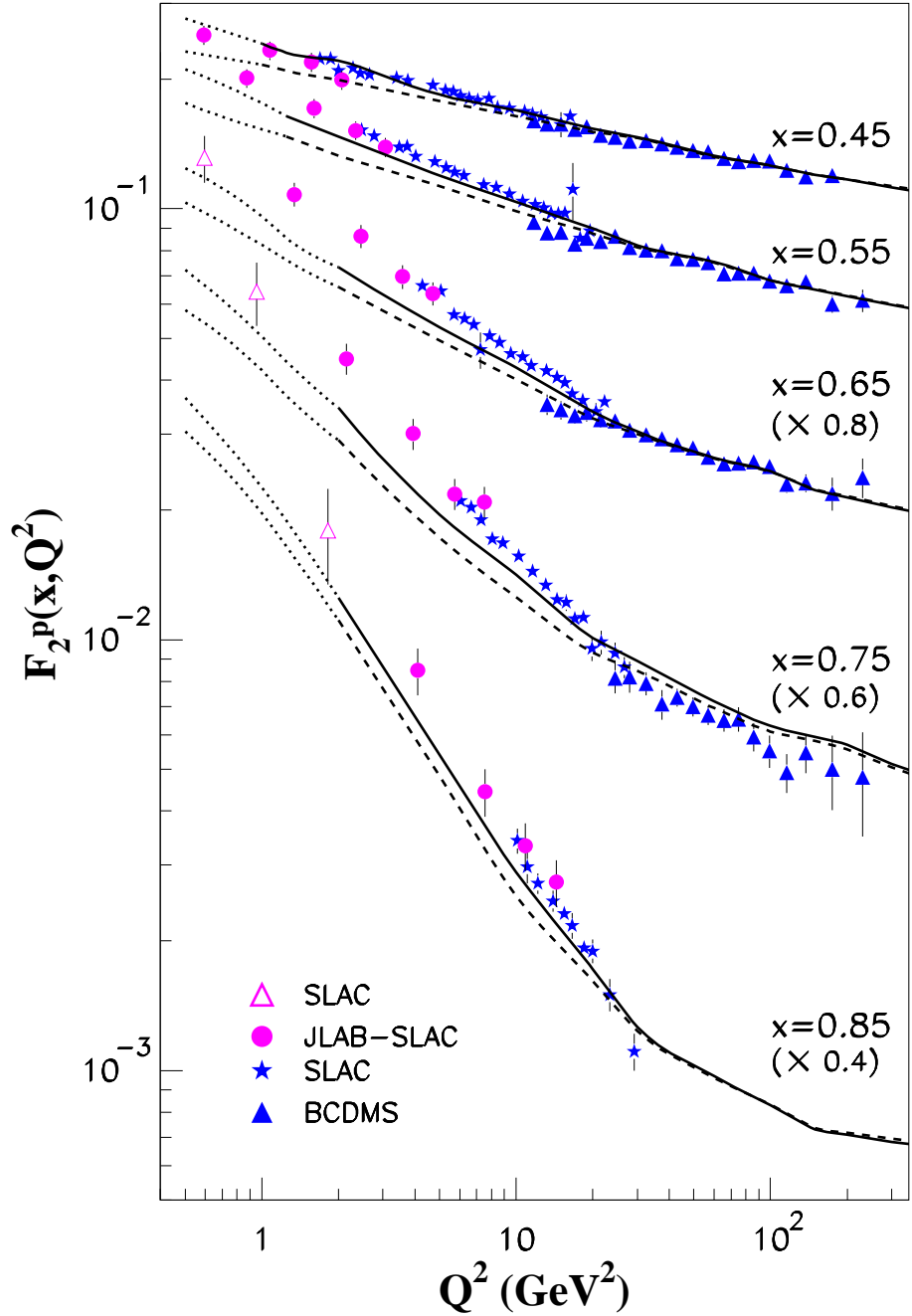}
\caption{Comparison of pQCD+TMC calculations at NLO (dashed lines)
and with resummation (full lines), with current large $x$ data. The full dots 
are in the resonance region, $1.3 \leq W^2 \leq 3.4$ GeV$^2$; the open triangles 
correspond to $W^2 \leq 1.3$ GeV$^2$. The dotted lines represent the regions 
where TMC contributions are uncertain.}
\end{figure}
\begin{figure}
\includegraphics[width=15cm]{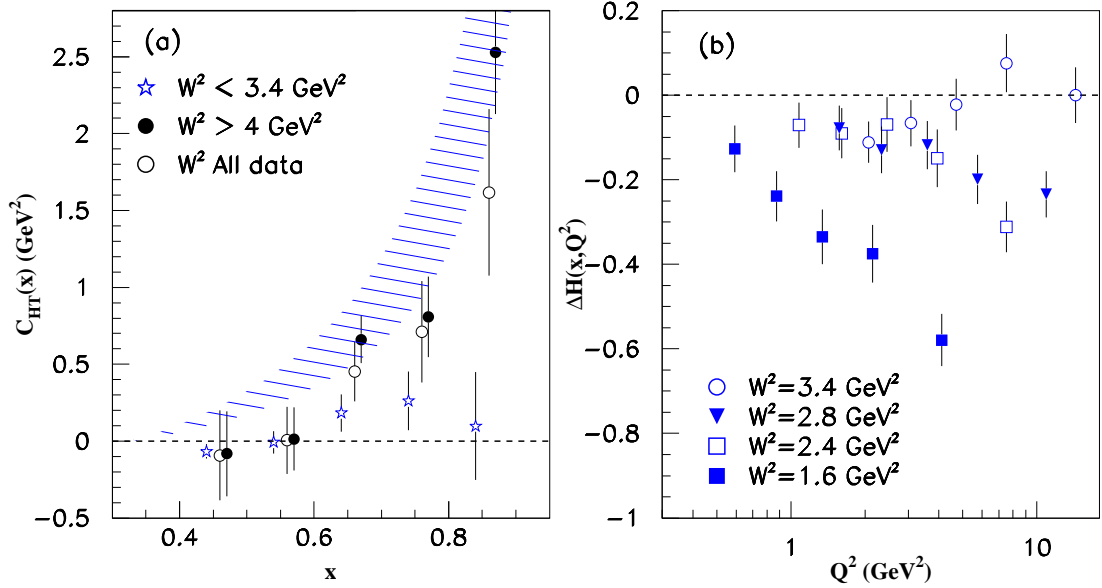}
\caption{(a) Coefficient $C_{HT}$, Eq.(\ref{CHT}), extracted from DIS data with
$W^2 \ge 4$ GeV$^2$ (full dots), from the resonance region, $W^2 < 4$ GeV$^2$
(stars) and averaged over the entire range of $W^2$ (open dots).     
The shaded area summarizes extractions previous to the current one. 
A dotted line at zero is added to guide the eye; 
(b) $\Delta H$, Eq.(\ref{DH}), extracted at fixed values of $W^2$ as described in 
the text, and plotted vs. $Q^2$. The figure further elucidates a breakdown of the
twist expansion at low $W^2$, already visible in (a).} 

%
\end{figure}
\end{document}